\documentclass[10pt,letterpaper]{llncs}
\usepackage{amsmath,amssymb,url,ifthen,fullpage}

\newcommand{\A}{\mathcal{A}}
\newcommand{\B}{\mathcal{B}}

\newcommand{\G}{\mathbb{G}}
\newcommand{\Z}{\mathbb{Z}}
\newcommand{\m}{\mathsf{m}}
\newcommand{\N}{\mathbb{N}}

\def \sample { \stackrel{\begin{footnotesize} _R
\end{footnotesize}}{\leftarrow} }

\title{Multi-Use Unidirectional Proxy Re-Signatures}

\author{Beno\^it Libert\inst{1} \and Damien Vergnaud\inst{2}}
\institute{
Universit\'e Catholique de Louvain, Crypto Group \\
Place du Levant, 3 -- 1348 Louvain-la-Neuve --
Belgium \\ \and
Ecole Normale Sup\'erieure -- C.N.R.S. -- I.N.R.I.A.  \\
45, Rue d'Ulm --  75230 Paris CEDEX 05 -- France  }

\begin{document}

\maketitle

\begin{abstract}
In 1998, Blaze, Bleumer, and Strauss suggested a cryptographic
primitive
  named \emph{proxy re-signatures}     where a proxy turns a
  signature computed under Alice's secret key into one from Bob on the same
  message. The semi-trusted proxy  does not learn either party's signing key
  and cannot sign arbitrary messages on behalf of  Alice or Bob. At
  CCS 2005, Ateniese and Hohenberger revisited the primitive by
  providing
  appropriate security definitions and efficient constructions in the
  random oracle model. Nonetheless, they left open the problem of
   designing  a {\it multi-use unidirectional} scheme where the proxy is  able
  to translate in only one direction and signatures can be re-translated several times. \\
  \indent This paper solves this  problem, suggested for the first time $10$ years ago, and shows the first
  \emph{multi-hop}
  \emph{unidirectional} proxy re-signature schemes. We   describe  a random-oracle-using
  system that  is secure in the Ateniese-Hohenberger
  model.
  The same technique also yields  a similar construction in the {\it
  standard model} (i.e.  without relying on
  random oracles).
   Both
  schemes are efficient and require newly defined -- but falsifiable -- Diffie-Hellman-like assumptions
  in bilinear groups.

\medskip
\textbf{Keywords. } Multi-use proxy re-signatures,
unidirectionality,
    pairings.
\end{abstract}

\section{Introduction}
In 1998, Blaze, Bleumer  and Strauss \cite{BBS98} proposed a
cryptographic
  primitive  where a semi-trusted proxy is given some information
  that
  allows   turning \mbox{Alice}'s signature on a   message into Bob's
  signature on the same message. These \emph{proxy re-signatures}
   (PRS) -- not to be confused with proxy signatures \cite{MUO96} --  require that the proxy
   be unable to
  sign on behalf of
    Alice or Bob on its own. The last few years saw a renewed
    interest   in proxy re-cryptography
    \cite{AFGH05,AFGH05bis,AH05,GA,Hoh06,HRSV07,CH07}. \\
 \indent
This paper presents the first constructions of \emph{multi-use}
\emph{unidirectional} proxy re-signature wherein the proxy can only
translate signatures in one direction and messages can be re-signed
a polynomial number of times.  Our constructions are efficient  and
demand new (but falsifiable) Diffie-Hellman-related intractability
assumptions in bilinear map groups. One of our contributions is a
secure scheme in the standard   model (\emph{i.e.} without resorting
to the random oracle model).

\medskip
\noindent\textsc{Related work.}  Alice -- the delegator -- can
easily designate a proxy translating   signatures computed using
Bob's secret key  -- the delegatee -- into one that are valid w.r.t.
her public key by storing her secret key at the proxy. Upon
receiving Bob's signatures, the proxy can check them and re-sign the
message using Alice's private key. The problem with this approach is
that the proxy can sign arbitrary messages on behalf of Alice. Proxy
re-signatures aim at securely enabling the delegation of signatures
without fully trusting the proxy. They are related to proxy
signatures, introduced in  \cite{MUO96} and revisted in
\cite{DI,BPW03,MOY04}, in that any PRS can be used to implement a
proxy signature mechanism but
the converse is not necessarily true. \\
\indent In 1998, Blaze \emph{et al.}  \cite{BBS98} gave the first
example of PRS  where signing keys remain hidden from the proxy. The
primitive was formalized in 2005 by Ateniese and Hohenberger
\cite{AH05}  who pinned down useful properties that can be expected
from proxy re-signature schemes. \\
\begin{figure}[!h]
\begin{enumerate}
\item \textsf{Unidirectional:} re-signature keys can only be used for
  delegation in one direction;
\item \textsf{Multi-use:} a message can be re-signed a polynomial number of
  times;
\item \textsf{Private Proxy:} re-signature keys can be kept secret by
  an honest proxy;
\item \textsf{Transparent:} a user may not even know that a proxy exists;
\item \textsf{Unlinkable:} a re-signature cannot be linked to the one
  from which it was generated;
\item \textsf{Key optimal:} a user is only required to store a
  constant amount of secret data;
\item \textsf{Non-interactive:} the delegatee does not act in the
  delegation process;
\item \textsf{Non-transitive:} the proxy cannot re-delegate signing rights;
\end{enumerate}
\label{fig:properties}\vspace{-0.8 cm}
\end{figure}
\indent Blaze \emph{et al.}'s construction is \emph{bidirectional}
(\emph{i.e.} the
  proxy information allows  ``translating'' signatures in either direction) and \emph{multi-use} (\emph{i.e.} the
  translation of signatures can be performed in sequence and multiple
  times by distinct proxies without requiring the intervention of
  signing entities). Unfortunately, Ateniese and Hohenberger \cite{AH05}
  pinpointed a flaw in the latter scheme: given a
  signature/re-signature pair, anyone can deduce the re-signature key that has
  been used in the delegation  (\emph{i.e.}
  the \emph{private proxy} property is not satisfied). Another issue in \cite{BBS98} is that
  the proxy and the delegatee can collude to expose the delegator's
  secret. \\
  \indent
To overcome these limitations, Ateniese and Hohenberger proposed two
constructions based on bilinear maps. The first one
is
a quite simple multi-use, bidirectional protocol built on
Boneh-Lynn-Shacham (BLS) signatures \cite{BLS}. Their second scheme
is unidirectional (the design of such a scheme was an
open problem raised in \cite{BBS98}) but single-use.
It involves two different signature algorithms: \emph{first-level}
signatures   can  be translated by the proxy whilst
\emph{second-level} signatures   cannot. A slightly less efficient
variant
was also suggested to ensure the privacy of re-signature keys kept
at the proxy. The security of all schemes was analyzed in the random
oracle model \cite{BR}. \\ \vspace{-0.3 cm}

\noindent\textsc{Our contributions.} Ateniese and Hohenberger left
as  open challenges the design of  multi-use  unidirectional
 systems and that of secure schemes in the
standard security model. The present paper solves both problems:
\begin{itemize}
\item we present a simple and efficient system (built  on the short signature put forth by Boneh \emph{et
al.} \cite{BLS}) which is secure in the random oracle
  model under a reasonable extension of the Diffie-Hellman  assumption;
\item using an elegant technique due to Waters \cite{Wat05},
    the scheme is easily modified so as to achieve security in the standard
  model. To the best of our knowledge, this actually provides the
  first unidirectional PRS that dispenses with random oracles
    and thereby improves a recent  bidirectional construction
  \cite{SCWL}.
\end{itemize}
 Both
proposals additionally preserve the privacy of proxy keys (with an
improved efficiency w.r.t. \cite{AH05} in the case of the first
one). They  combine almost all of the above properties. As  in
prior unidirectional schemes, proxies are not completely transparent
since signatures have different shapes and lengths  across successive levels.
 The size of our signatures actually grows linearly
with the number of past translations: signatures at level $\ell$
(i.e. that have been translated $\ell-i$ times if the original
version was signed at level $i$) consist of about $2\ell$ group
elements. In spite of this blow-up, we retain  important benefits:
\begin{itemize}
\item signers may want to tolerate a limited number (say $t$) of signature translations for specific messages.
Then, if at most $L$  translations are permitted in the global system,
users can directly generate a signature at level $L-t$.
\item the conversion of a $\ell^{\textrm{th}}$ level signature
is indistinguishable from one generated  at level $\ell+1$ by the
second signer. The original signer's identity is moreover
perfectly hidden and the verifier only needs the new signer's
public key.
\end{itemize}
\indent  The simplicity of our schemes makes them attractive for
applications that motivated the search for multi-use unidirectional
systems in \cite{AH05}. One of them was to provide a proof that a
certain path was taken in a directed graph: for instance, U.S.
customs only need one public key (the one of the immigration agent
who previously validated a signature on an e-passport) to make sure
that a foreign visitor legally entered the country and went through
the required checkpoints. Another application was the conversion of
certificates where valid signatures for untrusted public keys can be
turned into signatures that verify under trusted keys. As
exemplified in \cite{AH05}, unidirectional schemes are quite
appealing for converting certificates between {\it ad-hoc} networks:
using the public key of network B's certification authority (CA),
the CA of network A can non-interactively compute a translation key
and set up a proxy converting certificates from network B within its
own domain without having to rely on untrusted nodes of B. \\
\vspace{-0.3 cm}



\noindent\textsc{Roadmap.} In the forthcoming sections, we recall
the syntax of  unidirectional PRS schemes and the security model
in section \ref{sec:model}.  Section
 \ref{sec:bilinear} explains which algorithmic assumptions we need. Section \ref{BLS-Version}    describes our
random-oracle-using scheme.
 Section \ref{Waters-version} details how to get rid of the random
oracle idealization.

\vspace{-0.2 cm}

\section{Model and Security Notions}\label{sec:model}

We first recall the syntactic definition of unidirectional PRS
schemes from \cite{AH05}.
\begin{definition}[Proxy Re-Signatures]
A (unidirectional) proxy re-signature (PRS) scheme for $N$ signers
and $L$ levels (where $N$ and $L$ are both polynomial in the
security parameter $\lambda$) consists of a tuple of (possibly
randomized)
  algorithms
$(\mathsf{Global\textrm{-}Setup},\mathsf{Keygen},\mathsf{ReKeygen},\mathsf{Sign},\mathsf{Re\textrm{-}Sign},\mathsf{Verify})$
where:
\begin{description}
 \item[$\mathsf{Global\textrm{-}Setup}(\lambda)$:] is a randomized algorithm (possibly run by a trusted party) that takes as   input a
 security parameter $\lambda$ and produces  a set of system-wide public
 parameters $\mathsf{cp}$.
 \item[$\mathsf{Keygen}(\mathsf{cp})$:] is a probabilistic algorithm that, on
 input of public parameters $\mathsf{cp}$, outputs  a signer's
 private/public key pair $(sk,pk)$.
 \item[$\mathsf{ReKeygen}(\mathsf{cp},pk_i,sk_j)$:] on input of
 public parameters $ \mathsf{cp}$, signer $i$'s public key $pk_i$
 and signer $j$'s private key $sk_j$, this (ideally non-interactive) algorithm outputs a
 re-signature key $R_{ij}$ that allows translating $i$'s
 signatures into  signatures in the name of $j$.
 \item[$\mathsf{Sign}(\mathsf{cp},\ell,sk_i,m)$:] on input of
 public parameters $\mathsf{cp}$, a message $m$, a private key
 $sk_i$ and an integer
 $\ell \in \{1,\ldots,L\}$, this (possibly probabilistic)
 algorithm outputs a signature $\sigma$ on behalf of signer $i$ at
 level $\ell$.
  \item[$\mathsf{Re\textrm{-}Sign}(\mathsf{cp},\ell,m,\sigma,R_{ij},pk_i,pk_j)$:]
  given common   parameters $\mathsf{cp}$,  a  level $\ell< L$ signature $\sigma$ from signer
  $i\in \{1,\ldots,N\}$ and a re-signature key $R_{ij}$, this
  (possibly randomized) algorithm first checks that  $\sigma$ is valid w.r.t $pk_i$. If yes, it outputs a
  signature $\sigma'$ which verifies at level  $\ell+1$   under public key $pk_j$.
 \item[$\mathsf{Verify}(\mathsf{cp},\ell,m,\sigma,pk_i )$:]  given public parameters
 $\mathsf{cp}$, an integer $\ell \in \{1,\ldots,L\}$, a message
 $m$, an alleged signature $\sigma$ and a public key $pk_i$, this deterministic
 algorithm outputs $0$ or $1$.
\end{description} For all security parameters $\lambda \in \N$ and
 system-wide   parameters $\mathsf{cp}$ output by
 $\mathsf{Global\textrm{-}Setup}(\lambda)$, for all couples of
  private/public key pairs $(sk_i,pk_i)$, $(sk_j,pk_j)$ produced by
 $\mathsf{Keygen}(\mathsf{cp})$, for any $\ell \in \{1,\ldots,L\}$ and
  message $m$, we should have
\begin{eqnarray*}
&
\mathsf{Verify}(\mathsf{cp},\ell,m,\mathsf{Sign}(\mathsf{cp},\ell,sk_i,m),pk_i
)= 1; & \\
&\mathsf{Verify}(\mathsf{cp},\ell,m,\mathsf{ReSign}(\mathsf{cp},\ell,m,\mathsf{Sign}(\mathsf{cp},\ell,sk_i,m),\mathsf{ReKeygen}(\mathsf{cp},pk_i,sk_j)),pk_j
) = 1.&\\
\end{eqnarray*}
\end{definition}
To lighten notations, we  sometimes omit to explicitly include
public parameters $\mathsf{cp}$ that are part of the input of all
but one algorithms. \\
\indent The security model of \cite{AH05} considers the following
two orthogonal notions termed \emph{external} and \emph{insider
security}.

\begin{description}
\item[External security:] is the security against adversaries
outside the system (that differ from the proxy and delegation
partners). This notion demands that the next
probability be a negligible   function of the security parameter
$\lambda$:
\begin{eqnarray*}
\textrm{Pr}[ \{pk_i,sk_i) & \leftarrow &  \mathsf{Keygen}(\lambda)
\}_{i\in [1,N]},\\ && (i^\star,L,m^\star,\sigma^\star) \leftarrow
\A^{\mathcal{O}_{Sign}(.),\mathcal{O}_{Resign}(.)}(\{pk_i\}_{i\in
[1,N]}): \\ & & \mathsf{Verify}(L,pk_{i^\star},m^\star,\sigma^\star)
\wedge (i^\star,m^\star) \not\in Q ]
\end{eqnarray*}
where $\mathcal{O}_{Sign}(.)$ is an oracle taking as input a
message and an index  $i \in \{1,\ldots,N\}$  to return a first
level signature $\sigma \leftarrow \mathsf{Sign}(1,sk_i,m)$; the
oracle $\mathcal{O}_{Resign}(.)$ takes as input indices $i,j \in
\{1,\ldots,N\}$ and a level $\ell$ signature $\sigma$ and returns
the output of $\sigma' \leftarrow
\mathsf{Re\textrm{-}Sign}(\ell,m,\sigma,\mathsf{ReKeygen}(pk_i,sk_j))$;
and $Q$ denotes the set of (signer,message) pairs $(i,m)$ queried
to $\mathcal{O}_{Sign}(.)$ or such that a tuple $(?,j,i,m)$, with
$j \in \{1,\ldots,N\}$, was queried to $\mathcal{O}_{Resign}(.)$.
 This notion only
makes sense if re-signing keys are kept private by the proxy. \\
\vspace{-0.3 cm}
\item[Internal security:] The second security notion considered in
\cite{AH05} strives to protect users, as much as possible, against
dishonest proxies and colluding delegation partners. Three security
guarantees should be ensured. \\ \vspace{-0.3 cm}
\begin{itemize}
\item[1.] {\bf Limited Proxy security:} this notion captures the
proxy's inability to sign messages on behalf of the delegatee or to
create
 signatures for the delegator unless messages were   first signed by
one of the latter's delegatees.  Formally, we consider a game where
adversaries have all re-signing keys but are denied access to
signers' private keys. The following probability should be
negligible:
\begin{eqnarray*}
\textrm{Pr}\big[ \{pk_i,sk_i)  \leftarrow  \mathsf{Keygen}(\lambda)
\}_{i\in [1,N]},~ \{R_{ij}   \leftarrow  \mathsf{ReKeygen}(pk_i,sk_j)
\}_{i,j \in [1,N]},  & &     \\  (i^\star,L,m^\star,\sigma^\star) \leftarrow
\A^{\mathcal{O}_{Sign}(.,.)}
\big(\{pk_i\}_{i\in [1,N]},  \{R_{ij}
\}_{i,j \in [1,N]}  \big): & & \\
\mathsf{Verify}(L,pk_{i^\star},m^\star,\sigma^\star) \wedge m^\star
\not\in Q \big]
\end{eqnarray*}
where $\mathcal{O}_{Sign}(.,.)$ is an oracle taking as input a
message and an index $i \in \{1,\ldots,N\}$ to return a first level
signature $\sigma \leftarrow \mathsf{Sign}(1,sk_i,m)$   and $Q$ stands for the set of
messages $m$ queried to the signing oracle. \\ \vspace{-0.3 cm}
\item[2.] {\bf Delegatee Security:} informally, this notion
protects the delegatee from a colluding delegator and proxy. Namely,
the delegatee is assigned the index $0$. The adversary is provided
with an oracle returning first level signatures on behalf of $0$ and
 is also granted access to re-signature keys\footnote{In non-interactive schemes, the adversary can compute those keys herself from $pk_0$ and $sk_i$,
 with $i\neq 0$,  and the definition
can be   simplified. In the general case, they remain part of the adversary's input.} $R_{0i}$ for all $i\neq 0$
(but not $R_{i0}$ for any $i$). Her probability of success
\begin{eqnarray*}
\textrm{Pr}\big[&&\{pk_i,sk_i)  \leftarrow  \mathsf{Keygen}(\lambda)
\}_{i\in [0,N]},   \\ & & \{R_{ij}   \leftarrow  \mathsf{ReKeygen}(pk_i,sk_j)
\}_{i \in  \{0,\ldots,N\},j \in \{1,\ldots,N\}}  \\ & &  (L,m^\star,\sigma^\star) \leftarrow
\A^{\mathcal{O}_{Sign}(0,.)}
\big(pk_0,\{pk_i,sk_i\}_{i\in [1,N]},\{R_{ij}
\}_{i \in  \{0,\ldots,N\},j \in \{1,\ldots,N\}} \big):  \\
& &  \mathsf{Verify}(L,pk_{0},m^\star,\sigma^\star) \wedge m^\star
\not\in Q ~\big],
\end{eqnarray*}
where  $Q$ is the set of messages queried to $\mathcal{O}_{Sign}(0,.)$, should be negligible. \\ \vspace{-0.3 cm}
\item[3.] {\bf Delegator Security:} this notion captures that a
collusion between the delegatee and the proxy should be harmless for
the honest delegator.  Namely, we consider a target delegator with
index $0$. The adversary is given private keys of all other signers
$i \in \{1,\ldots,N\}$ as well as {\it all} re-signature keys
including $R_{i0}$ and $R_{0i}$ for $i \in \{1,\ldots,N\}$. A
signing oracle $\mathcal{O}_{Sign}(0,.)$ also provides her with
first level signatures for $0$. Yet, the following  probability
should be negligible,
\begin{eqnarray*}
\textrm{Pr}\big[ \{pk_i,sk_i)  \leftarrow  \mathsf{Keygen}(\lambda)
\}_{i\in [0,N]},  \{R_{ij} \leftarrow  \mathsf{ReKeygen}(pk_i,sk_j)
\}_{i,j \in [0,N]}, & & \\  (1,m^\star,\sigma^\star) \leftarrow
\A^{\mathcal{O}_{Sign}(0,.)}
\big(pk_0,\{pk_i,sk_i\}_{i\in [1,N]},  \{R_{ij}\}_{i,j \in [0,N]},   \big): & & \\
\mathsf{Verify}(1,pk_{0},m^\star,\sigma^\star) \wedge m^\star
\not\in Q \big],
\end{eqnarray*}
  meaning   she has little chance of framing user $0$ at the first level.
\end{itemize}
\end{description}
 An important difference between   external   and   limited proxy
 security should be underlined. In the former, the attacker is
 allowed to obtain signatures on the target message $m^\star$ for
  signers other than $i^\star$. In the latter, the target message
 cannot be queried for signature at all (knowing   all proxy keys, the attacker would trivially win
 the game otherwise).

\section{Bilinear Maps and Complexity Assumptions}\label{sec:bilinear}

\noindent\textsc{Bilinear  groups.} Groups $(\G,\G_T)$ of prime
order $p$ are called \emph{bilinear map groups} if there is a
mapping $e: \G \times \G \rightarrow \G_T$ with the following
properties:
\begin{enumerate}
\item[1.] bilinearity: $e(g^a,h^b)=e(g,h)^{ab}$ for any $(g,h)\in \G\times \G$ and $a,b\in \mathbb{Z}$;
\item[2.] efficient computability for any  input pair;
\item[3.] non-degeneracy: $e(g,h)\neq 1_{\G_T}$ whenever $g,h\neq 1_{\G}$.
\end{enumerate}
\indent In these groups, we  assume the hardness of the well-known
Computational Diffie-Hellman (CDH) problem which is to compute
$g^{xy}$ given $g^x$ and $g^y$.

\medskip
\noindent\textsc{Flexible Diffie-Hellman problems.} Our signatures
rely on new   generalizations of the
  Diffie-Hellman problem. To motivate  them, let us
first recall the definition of the \emph{2-out-of-3 Diffie-Hellman}
problem \cite{KJP06}.

\begin{definition}
In a prime order group $\G$, the {\bf
$\mathbf{2}$-out-of-$\mathbf{3}$ Diffie-Hellman } problem
($2$-$3$-CDH) is, given $(g,g^a,g^b)$, to find a pair $(C,C^{ab})
\in \G \times \G$ with $C \neq 1_{\G}$.
\end{definition}
We introduce a potentially harder version of this problem that we
call $1$-Flexible Diffie-Hellman problem:

\begin{definition}
The {\bf $\mathbf{1}$-Flexible Diffie-Hellman} problem ($1$-FlexDH)
is, given $(g,A=g^a,B=g^b) \in \G^3$, to find a triple
$(C,C^a,C^{ab}) \in (\G\backslash\{1_{\G}\})^3$.
\end{definition}
The unforgeability of our multi-use unidirectional proxy
re-signatures is proved assuming the intractability of a relaxed
variant of this problem where   more flexibility is permitted in the
choice of the base $C$ for the Diffie-Hellman computation.
\begin{definition}
The {\bf $\mathbf{\ell}$-Flexible Diffie-Hellman} problem
($\ell$-FlexDH) is, given $(g,A=g^a,B=g^b) \in \G^3$, to find a
$(2\ell+1)$-uple
$$(C_1,\ldots,C_\ell,D_1^a,\ldots,D_\ell^a,D_{\ell}^{ab}) \in \G^{2\ell+1}   $$
where $\log_g(D_j) = \prod_{i=1}^j \log_g(C_i) \neq 0$   for $j \in
\{1,\ldots\ell\}$.
\end{definition}
A given  instance  has many publicly verifiable solutions: a
candidate $2\ell+1$-tuple
$(C_1,\ldots,C_\ell,D_1',\ldots,D_\ell',T)$ is  acceptable if
$e(C_1,A)=e(D_1',g)$, $e(D_j',g)=e(D_{j-1}',C_j)$ for
$j=2,\ldots,\ell$ and $e(D_{\ell}',B)=e(T,g)$. The $\ell$-FlexDH
assumption is thus  falsifiable according to  Naor's classification
 \cite{Naor}.

 In generic groups, the general intractability result given by
   theorem $1$ of \cite{KJP06} by Kunz-Jacques and Pointcheval implies the generic hardness of
$\ell$-FlexDH. For completeness, appendix \ref{generic} gives an
adaptation of this result   in generic \emph{bilinear} groups.

\begin{remark}
The \emph{knowledge-of-exponent assumption} (KEA1) \cite{BP04} was
introduced in 1991 by Damg\aa{}rd \cite{Dam91}. Roughly speaking,
KEA1 captures the intuition  that any algorithm which, given elements $(g,g^x)\in\G^2$, computes a pair $(h,h^x)\in\G^2$ must
``know'' $\log_{g}(h)$. Under KEA1, the intractability of the
$\ell$-Flexible Diffie-Hellman problem  is easily seen to be boil
down to the Diffie-Hellman assumption. Given $(g,g^a)$, an
adversary outputting $(C_1,D_1^a)=(C_1,C_1^a)$ necessarily
``knows'' $t_1=\log_g C_1$ and thus also
$(C_2,C_2^a)=(C_2,(D_2^a)^{1/t_1})$ as well as $t_2=\log_g C_2$,
which in turn successively yields logarithms of
$C_3,\ldots,C_{\ell}$. Although the KEA1 assumption is inherently
non-falsifiable, it   holds in  generic groups \cite{Dent06,AF07}
and  our results can be seen as resting on the combination
CDH+KEA1.
\end{remark}

\medskip
\noindent\textsc{Modified Diffie-Hellman problem.} The second
assumption that we need is that the CDH problem $(g^a,g^b)$ remains
hard even when $g^{(a^2)}$ is available.
\begin{definition}
The {\bf modified  Computational Diffie-Hellman } problem (mCDH) is,
given $(g,g^a,g^{(a^2)},g^b)\in \G^4$, to compute $g^{ab} \in \G$.
\end{definition}
In fact, we
  use an  equivalent formulation of
the problem which is to find $h^{xy}$ given $(h,h^x,h^{1/x},h^{y})$
(the equivalence is readily observed by  defining $g=h^{1/x}$,
$x=a$, $y=b/a$).

\section{A Multi-Hop Scheme in the Random Oracle
Model} \label{BLS-Version}

   To provide a
better intuition of the underlying idea of our scheme, we first
describe its single-hop version before extending it into a
multi-hop
system. \\
\indent Our approach slightly differs from the one in \cite{AH05}
where signers have a ``strong'' secret and a ``weak'' secret that
are respectively used to produce  first and second level signatures.
In our scheme, users have a single secret but first and second level
signatures retain different shapes. Another
difference is that our re-signature algorithm is probabilistic. \\
\indent   We exploit the idea that, given $g^b  \in \G=\langle g
\rangle$ for some $b\in \Z$, one can hardly generate a
Diffie-Hellman triple $(g^a,g^b,g^{ab})$ without knowing the
corresponding exponent $a$ \cite{Dam91}. A valid BLS signature
\cite{BLS} $(\sigma=H(m)^x,X=g^x)$  can be blinded into
$(\sigma'_1,\sigma_2')=(\sigma^t,X^t)$ using a random exponent $t$.
An extra  element $g^t$   then serves as evidence that
$(\sigma'_1,\sigma_2')$ actually hides a valid pair. This technique
can be iterated several times  by adding two group elements at each
step.   To translate signatures from signer $i$ to signer $j$, the
key idea is to have  the proxy perform an appropriate change of
variable involving the
 translation key during the blinding.

 \indent The scheme is obviously not strongly
unforgeable in the sense of \cite{ADR} (since all but first level
signatures can be publicly re-randomized) but this ``malleability''
of signatures is not a weakness whatsoever. It  even turns out to be
a desirable feature allowing for the unlinkability of translated
signatures w.r.t. original ones.  \\ \vspace{-0.4 cm}

\subsection{The Single Hop Version}

In this scheme, signers' public keys consist of a single group
element $X=g^x \in \G$. Their well-formedness is thus efficiently
verifiable by the certification authority that just has to check
their membership in $\G$. This already improves \cite{AH05} where
public keys $(X_1,X_2)=(g^x,h^{1/x}) \in \G^2$ ($g$ and $h$ being
common parameters) must be validated by testing whether
$e(X_1,X_2)=e(g,h)$.
\begin{description}
\item[\textsf{Global-setup}$(\lambda)$:]   this algorithm
chooses bilinear   groups $(\G,\G_T)$ of prime order
$p>2^{\lambda}$. A generator $g \in \G$ and a hash function
$H:\{0,1\}^* \rightarrow \G$ (modeled as a random oracle in the
security proof) are also chosen. Public parameters only consist of
$\mathsf{cp}:=\{\G,\G_T,g,H \}.$
\item[\textsf{Keygen}$(\lambda)$:] user $i$'s public key is set as $X_i=g^{x_i}$ for a random $x_i \sample \Z_p^*$.
\item[\textsf{ReKeygen}$(x_j,X_i)$:]  this algorithm outputs the proxy key
$R_{ij}=X_i^{1/x_j}=g^{x_i/x_j}$  which allows
  turning signatures from $i$ into signatures from $j$. \\
\vspace{-0.3 cm}
\item[\textsf{Sign}$(1,x_i,m)$:] to sign   $m \in \{0,1\}^*$ at
level $1$, compute $\sigma^{(1)}=H(m)^{x_i} \in \G$. \\
\vspace{-0.3 cm}
\item[\textsf{Sign}$(2,x_i,m)$:] to sign $m \in \{0,1\}^*$ at level $2$,    choose  $t \sample \Z_p^*$ and
compute
 \begin{eqnarray} \label{yes-ca-marche-1}
 \sigma^{(2)}=(\sigma_0,\sigma_1,\sigma_2)=(H(m)^{x_i t}
,X_i^t,g^t).
 \end{eqnarray}
\item[\textsf{Re-Sign}$(1,m,\sigma^{(1)},R_{ij},X_i,X_j)$:] on input of   $m \in \{0,1\}^*$, the re-signature key
$R_{ij}=g^{x_i/x_j}$, a signature $\sigma^{(1)} \in \G$ and   public
keys  $X_i,X_j$, check the validity of $\sigma^{(1)}$ w.r.t signer
$i$ by testing $e(\sigma^{(1)},g)=e(H(m),X_i)$.
  If valid, $\sigma^{(1)}$ is turned into a signature on behalf of $j$ by choosing $ t \sample \Z_p^*$ and computing
\begin{eqnarray*}
 \sigma^{(2)}=(\sigma_0',\sigma_1',\sigma_2')=({\sigma^{(1)}}^{t}, X_i^t,R_{ij}^t)
  = ( H(m)^{x_it},X_i^t,g^{tx_i/x_j} )
\end{eqnarray*}
  If we set $\tilde{t}=tx_i/x_j$, we have
  \begin{eqnarray} \label{yes-ca-marche-2}
  \sigma^{(2)}=(\sigma_0',\sigma_1',\sigma_2')=(H(m)^{ x_j  \tilde{t} }
 ,X_j^{\tilde{t}},g^{\tilde{t}}).
 \end{eqnarray}
\item[\textsf{Verify}$(1,m,\sigma^{(1)},X_i)$:] this algorithm accepts if $e(\sigma^{(1)},g)=e(H(m),X_i)$.  \\
\vspace{-0.3 cm}
\item[\textsf{Verify}$(2,m,\sigma^{(2)},X_i)$:]   a second level signature
$\sigma^{(2)}=( \sigma_0,\sigma_1,\sigma_2)$ is  accepted for the
public key $X_i$ if the following conditions are true.
\begin{eqnarray*} \label{validity-21-BLS}
e(\sigma_0,g) = e(\sigma_1,H(m)) \qquad e(\sigma_1,g) =
e(X_i,\sigma_2)
\end{eqnarray*}

\end{description}
Relations (\ref{yes-ca-marche-1}) and (\ref{yes-ca-marche-2}) show
that translated signatures have exactly the same distribution  as
signatures directly produced by signers at level $2$. \\
\indent In comparison with the only known  unidirectional PRS with
private re-signing keys (suggested in section 3.4.2 of \cite{AH05}),
this one features shorter second level signatures that must include
a Schnorr-like \cite{Sch} proof of knowledge in addition to $3$
group elements in \cite{AH05}. On the other hand, signatures of
\cite{AH05} are strongly unforgeable unlike ours. \\
\indent It is also worth mentioning that the above scheme only
requires the $1$-Flexible Diffie-Hellman assumption which is more
classical than the general $\ell$-FlexDH.

\subsection{How to Obtain Multiple Hops} \label{sec:ellscheme}

The above construction can be scaled up into a multi-hop PRS
scheme if we iteratively apply the same idea several times. To
prevent the linkability of signatures between successive levels
$\ell+1$ and $\ell+2$, the re-signature algorithm performs a
re-randomization using random exponents $r_1,\ldots,r_{\ell}$.

\begin{description}
\item[\textsf{Sign}$(\ell+1,x_i,m)$:] to sign   $m \in \{0,1\}^*$ at  the
$(\ell+1)^{\textrm{th} }$ level,   user $i$ chooses
$(t_1,\ldots,t_{\ell}) \sample (\Z_p^*)^{\ell}$ and outputs
$\sigma^{(\ell+1)}=(\sigma_0,\ldots,\sigma_{2\ell}) \in
\G^{2\ell+1}$ where
\begin{eqnarray*}
\sigma_0 = H(m)^{x_i t_1 \cdots t_\ell}, \qquad
\left\{
\begin{array}{ll}
\sigma_k  =  g^{x_i t_1 \cdots t_{\ell+1-k}}  & \textrm{ for } k \in
\{1,\ldots, \ell\} \\
\sigma_k  =  g^{t_{k-\ell}} &  \textrm{ for } k \in
\{\ell+1,\ldots,2\ell\}.
\end{array}
   \right.
\end{eqnarray*}
\item[\textsf{Re-Sign}$(\ell+1,m,\sigma^{(\ell+1)},R_{ij},X_i,X_j)$:] on input of a message
  $m \in \{0,1\}^*$, the re-signature key $R_{ij}=g^{x_i/x_j}$, a
  valid
  $(\ell+1)^{\textrm{th}}$
  level signature
  \begin{eqnarray*}
  \sigma^{(\ell+1)} &=& (\sigma_0,\ldots,\sigma_{2\ell}) \\
  &=& (H(m)^{x_i t_1\cdots t_{\ell}}, g^{x_i t_1 \cdots t_{\ell}}, g^{x_i t_1 \cdots t_{\ell-1}}, \ldots,
  g^{x_it_1},g^{t_1},\ldots,g^{t_{\ell}} )  \in \G^{2\ell+1}
  \end{eqnarray*} and  public keys $X_i,X_j$,
  check the validity of $\sigma$ under $X_i$. If valid, $\sigma$  is turned into a $(\ell+2)^{\textrm{th}}$  level signature on behalf of $j$ by
  drawing $(r_0,r_1,\ldots,r_{\ell}) \sample (\Z_p^*)^{\ell+1}$ and
  computing $\sigma^{(\ell+2)}=(\sigma_0',\ldots,\sigma_{2\ell+2}') \in \G^{2\ell+3}$ where
\begin{eqnarray*}
\sigma_0'  =  \sigma_0^{r_0 \cdots r_{\ell}} ~\textrm{ and }~
\left\{
\begin{array}{ll}
\sigma_{k}' =  \sigma_k^{r_0 \cdots r_{\ell+1-k}} &  \textrm{ for } k \in
\{1,\ldots, \ell\} \\
\sigma_{\ell+1}'  =  X_i^{r_{0}}  & \\
\sigma_{\ell+2}'  =  R_{ij}^{r_{0}} & \\
\sigma_k'  =  \sigma_{k-2}^{r_{k-\ell-2}} &  \textrm{ for } k \in
\{\ell+3,\ldots,2\ell+2\}.\\
\end{array}
 \right.
\end{eqnarray*}
If we define $\tilde{t}_0=r_0x_i/x_j$ and $\tilde{t}_k=r_k t_k$
for $k=1,\ldots,\ell$, we  observe that
$$\sigma^{(\ell+2)} =(H(m)^{x_j \tilde{t}_0 \tilde{t}_1\cdots \tilde{t}_{\ell}}, g^{x_j \tilde{t}_0 \tilde{t}_1 \cdots \tilde{t}_{\ell}}, g^{x_j \tilde{t}_0 \tilde{t}_1 \cdots \tilde{t}_{\ell-1}}, \ldots,
  g^{x_j \tilde{t}_0},g^{\tilde{t}_0},\ldots,g^{\tilde{t}_{\ell}} )  \in \G^{2\ell+3} $$
\item[\textsf{Verify}$(\ell+1,m,\sigma^{(\ell+1)},X_i)$:]  the validity  of $\sigma^{(\ell+1)}=(\sigma_0,\ldots,\sigma_{2\ell}) \in
  \G^{2\ell+1}$ at level $(\ell+1)$  is checked by testing if these  equalities  simultaneously hold:
\begin{eqnarray*}
e(\sigma_0,g) & = & e(H(m),\sigma_1), \qquad e(\sigma_{\ell},g) = e(X_i,\sigma_{\ell+1}) \\
e(\sigma_k,g) & = & e(\sigma_{k+1},\sigma_{2\ell-k+1}) \hbox{ for
} k \in
 \{1,\ldots,\ell-1\}
\end{eqnarray*}

\end{description}
\vspace{-0.4 cm}

\subsection{Security}

\begin{theorem} \label{BLS-proof}
The $L$-level scheme  is a secure unidirectional proxy re-signature
under the $(L-1)$-FlexDH and mCDH assumptions in the random oracle
model.
\end{theorem}

\begin{proof}

\noindent\textit{Limited proxy security.} We show that an
adversary $\A_1$ with advantage $\varepsilon$ implies an algorithm
$\B_1$
 solving an $(L-1)$-FlexDH instance $(g,A=g^a,B=g^b)$ with
probability $O(\varepsilon/q_s)$, where $q_s$ is the number of
signing queries made by $\A_1$.

\begin{description}
\item[System parameters:]   $\A_1$ is challenged on parameters
  $\{\G,\G_T,g,\mathcal{O}_H\}$ where $\mathcal{O}_H$ is the
  random oracle controlled by the simulator $\mathcal{B}_1$.

\item[Public key generation:] when $\A_1$ asks for the creation of user $i
  \in \{1,\ldots,N\}$, $\B_1$ responds with a newly generated public key
  $X_i=A^{x_i}=g^{ax_i}$, for a random $x_i \sample \Z_p^*$, which
  virtually defines user $i$'s private key as $ax_i$.  For all pairs $(i,j)$,
  re-signature keys $R_{ij}$ are calculated as $R_{ij}=g^{x_i/x_j}=g^{ax_i/ax_j}$. \\ \vspace{-0.3 cm}
\item[Oracle queries:] $\A_1$'s queries are tackled with as follows.
  Following a well-known technique due to Coron \cite{Cor00}, a
  binary coin $c\in \{0,1\}$ with expected value $1-\zeta \in [0,1]$ decides whether
  $\B_1$ introduces the challenge in the output of the random
  oracle or an element of known signature. For the optimal value of
  $\zeta$, this introduces the  loss factor $O(q_s)$ in the success
  probability. \\ \vspace{-0.3 cm}
\begin{itemize}
\item[$\bullet$] {\it Random oracle queries}: To respond to these queries,
  $\B_1$ maintains a list (referred to as the $H$-List) of tuples
  $(m,h,\mu,c)$ as follows:
\begin{enumerate}
\item If the query $m$ already appears in the $H$-List, then $\B_1$ returns
  $h$;
\item Otherwise, $\B_1$ generates a random bit $c$ such that
  $\Pr[c=0] = \zeta$;
\item It picks uniformly at random $\mu \in \Z_p^*$ and computes $h=g^{\mu}$  if $c=0$ and
$h=B^{\mu}$ otherwise;
\item It adds the 4-uple $(m,h,\mu,c)$ to the $H$-List and returns $h$ as
  the answer to the random oracle query.
\end{enumerate}

\item[$\bullet$] {\it Signing queries}: when a signature of signer $i$
  is queried for a  message $m$, $\B_1$ runs the random oracle to obtain
 the 4-uple $(m,h,\mu,c)$ contained in the $H$-List. If $c=1$ then $\B_1$
  reports failure and aborts. Otherwise, the algorithm $\B_1$ returns
  $h^{x_i a} = A^{x_i \mu}$ as a valid signature on $m$. \vspace{-0.2 cm}
\end{itemize}
\end{description}
After a number of queries, $\A_1$ comes up with a message
$m^{\star}$, that was never queried for signature for any signer, an
index $i^\star \in \{1,\ldots,N\}$ and a $L^{\textrm{th}}$ level
forgery
${\sigma^\star}^{(L)}=({\sigma_0^\star},\ldots,{\sigma_{2L-2}}^\star)
\in \G^{2L-1}$. At this stage, $\B_1$ runs the random oracle to
obtain
 the 4-uple $(m^{\star},h^{\star},\mu^{\star},c^{\star})$ contained in the
   $H$-List and fails if $c^{\star} = 0$. Otherwise, if $ {\sigma^\star}^{(L)} $ is valid,
   it may be written
$$
({\sigma_0^\star},\ldots,{\sigma_{2L-2}}^\star)=\Big(B^{\mu^{\star}
x_{i^\star}
 at_1 \ldots t_{L-1}},A^{t_1,
     \ldots t_{L-1}}, \ldots, A^{t_1}, g^{t_1}, \ldots, g^{t_{L-1}} \Big)
$$
which provides $\B_1$ with a valid tuple
$(C_1,\ldots,C_{L-1},D_1^a,\ldots,D_{L-1}^a,D_{L-1}^{ab}) $, where
$D_{L-1}^{ab}={\sigma_0^\star}^{1/\mu^\star x_{i^\star}}$, so that
$\log_g(D_j) = \prod_{i=1}^j \log_g(C_i)$ for $j \in \{1,\ldots,
L-1\}$. A
  similar analysis to \cite{Cor00,BLS} gives the announced bound on
 $\B_1$'s advantage if the optimal probability $\zeta =q_s/(q_s+1)$ is used when answering hash
 queries. \\ \vspace{-0.3 cm}

\paragraph{Delegatee security.} We also show how to break the $(L-1)$-FlexDH assumption  out of a delegatee security adversary $\A_2$.
  Given an input pair $(A=g^a,B=g^b)$, the simulator
$\B_2$   proceeds  as  $\B_1$ did in the proof of limited proxy
security. \vspace{-0.2 cm}

\begin{description}
\item[System parameters and public keys:]   the target delegatee's public key  is set  $X_0=A=g^a$.  For $i=1,\ldots,n$, other
public keys  are defined as $X_i=g^{x_i}$ for a random $x_i \sample
\Z_p^*$. To generate re-signature keys $R_{ij}$, $\B_2$ sets $R_{ij}=g^{x_i/x_j}$ when  $i,j \neq 0$ and $R_{0j}=A^{1/x_j}=g^{a/x_j}$ for $j=1,\ldots,n$.   \\
\vspace{-0.3 cm}
\item[Queries:] $\A_2$'s hash and signing queries are handled   exactly  as in the proof of limited proxy security. Namely,
 $\B_2$  fails if $\A_2$ asks for a signature on a message $m$ for
 which $H(m)=B^{\mu}$ and    responds  consistently otherwise.  \\ \vspace{-0.3 cm}
\end{description}
When $\A_2$  outputs her   forgery ${\sigma^\star}^{(L)}=
(\sigma_0^\star,\ldots,\sigma_{2L-2}^\star)$ at level $L$, $\B_2$ is
successful if $H(m^\star)=B^{\mu^\star}$, for some $\mu^\star  \in
\Z_p^*$, and extracts an admissible $(2L-1)$-uple  as done in the
proof of limited proxy security. \\ \vspace{-0.3 cm}

\paragraph{Delegator security.} This security property is proven under the mCDH assumption. Given an adversary
$\A_3$ with advantage $\varepsilon$, we outline an algorithm
$\B_3$ that has probability $O(\varepsilon/q_s)$ of finding
$g^{ab}$ given $(g,A=g^a,A'=g^{1/a},B=g^b)$.
\begin{description}
\item[Public key generation:] as previously, the target public key is defined as   $X_0=A=g^a$.  Remaining public keys
 are set as $X_i=g^{x_i}$ for a random $x_i
\sample \Z_p^*$ for $i=1,\ldots,n$. This time, $\A_3$ aims at producing a first level forgery and is granted {\it all} re-signature keys,
including $R_{0j}$ and $R_{j0}$. For indexes $(i,j)$ s.t. $i,j \neq 0$,
$\B_3$ sets $R_{ij}=g^{x_i/x_j}$. If $i=0$, it calculates
$R_{0j}=A^{1/x_j}=g^{a/x_j}$. If $j=0$ (and thus $i \neq 0$), $\B_3$
computes $R_{i0}={A'}^{x_i}=g^{x_i/a}$ to $\A_3$.
\end{description}
\indent Hash and signing queries  are dealt with exactly as for previous adversaries. Eventually, $\A_3$  produces a first level
forgery ${\sigma^\star}^{(1)}$ for a new message $m^\star$. Then,
$\B_3$ can extract $g^{ab}$ if $H(m)=(g^b)^{\mu^\star} $ for some
$\mu^\star \in \Z_p^*$, which occurs with probability $O(1/q_s)$
using Coron's technique \cite{Cor00}. Otherwise, $\B_3$ fails. \\
\vspace{-0.3 cm}

\paragraph{External security.} We finally show that an external
security adversary $\A_4$ also allows breaking the $(L-1)$-FlexDH
assumption almost exactly as in the proof of limited proxy security.
The simulator $\B_4$ is given an instance $(g,A=g^a,B=g^b)$.  As
previously, $\B_4$ must ``program'' the random oracle $H$ hoping
that its output will be $H(m^\star)=B^{\mu^\star}$ (where $\mu^\star
\in \Z_p^*$ is known) for the message $m^\star$ that the forgery
${\sigma^\star}^{(L)} $ pertains to. The difficulty is that $\B_4$
must also be able to answer signing queries made on $m^\star$ for
all but one signers. Therefore, $\B_4$ must guess which signer
$i^\star$ will be $\A_4$'s prey beforehand. At the outset of the
game, it thus chooses an index $i^\star \sample \{1,\ldots,N\}$.
Signer $i^\star$'s public key is set as $X_{i^\star}=A=g^a$. All
other signers $i \neq i^\star$ are assigned public keys
$X_i=g^{x_i}$ for which $\B_4$ knows the matching secret $x_i$ and
can thus always answer signing queries.
\\ \indent Hash queries and signing queries involving $i^\star$ are
handled as in the proof of limited proxy security. When faced with
a re-signing query from $i$ to $j$ for a  valid signature
$\sigma^{(\ell)}$ at level $\ell \in \{1,\ldots,L\}$, $\B_4$
ignores $\sigma^{(\ell)}$ and simulates a first level signature
for signer $j$. The resulting signature $ {\sigma'}^{(1)}$ is then
turned into a $(\ell+1)^{\textrm{th}}$ level signature and given
back to $\A_4$. A re-signing query thus triggers a signing query
that only causes failure if  $H(m)$ differs from $g^{\mu}$ for a
known $\mu \in \Z_p^*$.
\\ \indent When $\A_4$ forges a signature at level $L$, $\B_4$   successfully
extract a $(2L-1)$-Flexible Diffie-Hellman tuple (as $\B_1$ and
$\B_2$ did) if $H(m^\star)=(g^b)^{\mu^\star}$ and if it correctly
guessed the identity $i^\star$ of the target signer. If $\A_4$'s
advantage is $\varepsilon$, we find  $O(\varepsilon/(
N(q_s+q_{rs}+1)))$ as a lower bound on $\B_4$'s probability of
success,   $q_{s}$ and $q_{rs}$ being the number of signature and
re-signature queries respectively.
 \qed
\end{proof}

\vspace{-0.4 cm}

\section{Eliminating the Random Oracle} \label{Waters-version}

 Several extensions of BLS signatures   have a
standard model counterpart when  Waters' technique supersedes random
oracle manipulations (e.g. \cite{LOSSW}). Likewise, we can very
simply twist our method and achieve the first unidirectional PRS
scheme (even including single hop ones) that avoids the random
oracle model. {\it Mutatis mutandis}, the scheme is totally similar
to
our first construction and relies on the same assumptions. \\

\subsection{The Single Hop Variant}

As in \cite{Wat05}, $n$ denotes the length of messages to be signed.
Arbitrary long messages can be signed if we first apply a
collision-resistant hash function with $n$-bit outputs, in which
case $n$ is part of the security parameter. \\
\indent The scheme requires a trusted party to generate common
public parameters. However, this party can remain off-line after the
setup phase.

\begin{description}
\item[\textsf{Global-setup}$(\lambda,n)$:] given   security parameters $\lambda,n$, this algorithm
chooses bilinear   groups $(\G,\G_T)$ of   order $p>2^{\lambda}$,
generators $g,h \sample \G$ and a random $(n+1)$-vector
$\overline{u}=(u',u_1,\ldots,u_n) \sample \G^{n+1}$. The latter
  defines a     function  $F:\{0,1\}^n \rightarrow \G$
mapping $n$-bit strings $\mathsf{m}=m_1\ldots m_n$ (where $m_i\in
\{0,1\}$ for all $i\in \{0,1\}$) onto $F(\m)= u' \cdot \prod_{i=1}^n
u_i^{m_i}$. The  public parameters are
$$\mathsf{cp}:=\{\G,\G_T,g,h,\overline{u} \}. $$
\item[\textsf{Keygen}$(\lambda)$:] user $i$ sets his public key as $X_i=g^{x_i}$ for a random $x_i \sample \Z_p^*$.
\item[\textsf{ReKeygen}$(x_j,X_i)$:] given user $j$'s private key
$x_j$ and user $i$'s public key $X_i$, generate the  re-signature
key $R_{ij}=X_i^{1/x_j}=g^{x_i/x_j}$ that will be used to translate
signature from $i$ into signatures from $j$.
\item[\textsf{Sign}$(1,\m,x_i)$:] to sign a message $\m=m_1\ldots m_n \in \{0,1\}^n$ at  the first
level, the signer  picks $r  \sample \Z_p^*$ at random and computes
\begin{eqnarray*}
\sigma^{(1)}=(\sigma_0,\sigma_1)=(h^{x_i} \cdot F(\m)^r ,g^r)
\end{eqnarray*}
\item[\textsf{Sign}$(2,\m,x_i)$:] to generate a second level
signature on  $\m=m_1\ldots m_n \in \{0,1\}^n$, the signer  chooses
$r,t \sample \Z_p^*$ and computes
\begin{eqnarray} \label{ca-marche-0}
 \sigma^{(2)}=(\sigma_0,\sigma_1,\sigma_2,\sigma_3)=(h^{tx_i} \cdot F(\m)^r ,g^r,X_i^t,g^t)
\end{eqnarray}
\item[\textsf{Re-Sign}$(1,\m,\sigma^{(1)},R_{ij},X_i,X_j)$:] on input of a message $\m \in \{0,1\}^n$, the re-signature key
$R_{ij}=g^{x_i/x_j}$, a signature $\sigma^{(1)}=(
\sigma_0,\sigma_1)$ and   public keys  $X_i,X_j$, check the validity
of $\sigma$ w.r.t signer $i$ by testing
 if
\begin{eqnarray} \label{validity-1}
 e(\sigma_0,g)=e(X_i,h) \cdot e(F(\m),\sigma_1)
\end{eqnarray}
 If $\sigma^{(1)}$ is a valid, it can be turned into a signature on behalf of $j$ by choosing $r',t \sample \Z_p^*$ and computing
\begin{eqnarray*}
 \sigma^{(2)}=(\sigma_0',\sigma_1',\sigma_2',\sigma_3')&=&(\sigma_0^{t} \cdot F(\m)^{r'} ,\sigma_1^t \cdot
 g^{r'},X_i^t,R_{ij}^t)\\
  &=& (h^{tx_i} \cdot F(\m)^{r''},g^{r''},X_i^t,g^{tx_i/x_j} )
\end{eqnarray*}
where $r''=tr+r'$. If we set $\tilde{t}=tx_i/x_j$, we observe that
\begin{eqnarray} \label{ca-marche}
  \sigma^{(2)}=(\sigma_0',\sigma_1',\sigma_2',\sigma_3')=(h^{\tilde{t}x_j} \cdot F(\m)^{r''} ,g^{r''},X_j^{\tilde{t}},g^{\tilde{t}})
\end{eqnarray}
\item[\textsf{Verify}$(1,\m,\sigma^{(1)},X_i)$:] the validity of a first level signature $\sigma^{(1)}=(\sigma_1,\sigma_2)$  is
checked by testing if relation (\ref{validity-1}) holds.  \\
\vspace{-0.3 cm}
\item[\textsf{Verify}$(2,\m,\sigma^{(2)},X_i)$:]   a second level signature
$\sigma^{(2)}=( \sigma_0,\sigma_1,\sigma_2,\sigma_3)$ is  accepted
for the public key $X_i$ if the following conditions are true.
\begin{eqnarray} \label{validity-21}
e(\sigma_0,g) &=& e(\sigma_2,h) \cdot e(F(\m),\sigma_1') \\
e(\sigma_2,g) &=& e(X_i,\sigma_3)  \label{validity-22}
\end{eqnarray}

\end{description}
To the best of our knowledge, the above scheme is the first
unidirectional PRS in the standard model and   solves another
problem  left open in \cite{AH05} where all constructions require
the random oracle model. Like the scheme of section
\ref{BLS-Version}, this extension of Waters' signature \cite{Wat05}
is scalable into a multi-hop PRS.


\subsection{The Multi-Hop Extension}

At levels $\ell \geq 2$, algorithms $\mathsf{Sign}$,
$\mathsf{Re\textrm{-}Sign}$ and $\mathsf{Verify}$ are generalized as
follows. \\ \vspace{-0.3 cm}

\begin{description}
\item[\textsf{Sign}$(\ell+1,m,x_i)$:] to sign   $\m \in \{0,1\}^n$
at level $\ell+1$,   user $i$ picks $r \sample \Z_p^*$,
$(t_1,\ldots,t_{\ell}) \sample (\Z_p^*)^{\ell}$ and outputs
$\sigma^{(\ell+1)}=(\sigma_0,\ldots,\sigma_{2\ell+1}) \in
\G^{2\ell+2}$ where
\begin{eqnarray*}
\left\{
\begin{array}{ll}
\sigma_0  =  h^{x_i t_1 \cdots t_\ell} \cdot F(\m)^r  &  \\
 \sigma_1  =   g^r & \\
\sigma_k  =  g^{x_i t_1 \cdots t_{\ell+2-k}} &  \hbox{ for } k \in
\{2,\ldots, \ell+1 \}\\
\sigma_k  =  g^{t_{k-\ell-1}} & \hbox{ for } k \in
\{\ell+2,\ldots,2\ell+1\}.
\end{array}\right.
\end{eqnarray*}
\item[\textsf{Re-Sign}$(\ell+1,\m,\sigma^{(\ell+1)},R_{ij},X_i,X_j)$:] on input of a message
  $\m \in \{0,1\}^*$, the re-signature key $R_{ij}=g^{x_i/x_j}$, a
 purported
  $(\ell+1)^{\textrm{th}}$
  level signature
  \begin{eqnarray*}
  \sigma^{(\ell+1)} &=& (\sigma_0,\ldots,\sigma_{2\ell+1}) \\
  &=& (h^{x_i t_1 \cdots t_{\ell}} \cdot F(\m)^r,g^r, g^{x_i t_1 \cdots t_{\ell}}, g^{x_i t_1 \cdots t_{\ell-1}}, \ldots,
  g^{x_it_1},g^{t_1},\cdots,g^{t_{\ell}} )  \in \G^{2\ell+2}
  \end{eqnarray*} and public keys $X_i,X_j$,
  check the correctness of $\sigma^{(\ell+1)}$ under $X_i$. If valid, $\sigma^{(\ell+1)}$ is translated for  $X_j$ by
  sampling $r' \sample \Z_p^*$, $(r_0,r_1,\ldots,r_{\ell}) \sample (\Z_p^*)^{\ell+1}$ and
  setting $\sigma^{(\ell+2)}=(\sigma_0',\ldots,\sigma_{2\ell+3}') \in \G^{2\ell+4}$ where
\begin{eqnarray*}
\left\{
\begin{array}{ll}
 \sigma_0'  =  \sigma_0^{r_0 \cdots r_{\ell}} \cdot F(\m)^{r'} &
 \\
  \sigma_1'  =  \sigma_1^{r_0 \cdots r_{\ell}} \cdot g^{r'} & \\
 \sigma_{k}'  =  \sigma_k^{r_0 \cdots r_{\ell+2-k}} & \hbox{ for } k \in
 \{2,\ldots, \ell+1\}\\
 \sigma_{\ell+2}'  =  X_i^{r_{0}} & \\
 \sigma_{\ell+3}'  =  R_{ij}^{r_{0}}  & \\
 \sigma_k'  =  \sigma_{k-2}^{r_{k-\ell-3}}   & \hbox{ for } k \in
 \{\ell+4,\ldots,2\ell+3\}.\\
\end{array}\right.
\end{eqnarray*}
If we define $\tilde{t}_0=r_0x_i/x_j$, $r''=r_0   \cdots
r_{\ell}+r'$ and $\tilde{t}_k=r_k t_k$ for $k=1,\ldots,\ell$, we
observe that
$$\sigma^{(\ell+2)}  =(h^{x_j \tilde{t}_0 \tilde{t}_1\cdots \tilde{t}_{\ell}} \cdot F(\m)^{r''},g^{r''}, g^{x_j \tilde{t}_0 \tilde{t}_1 \cdots \tilde{t}_{\ell}}, g^{x_j \tilde{t}_0 \tilde{t}_1 \cdots \tilde{t}_{\ell-1}}, \ldots,
  g^{x_j \tilde{t}_0},g^{\tilde{t}_0},\ldots,g^{\tilde{t}_{\ell}} )  $$
\item[\textsf{Verify}$(\ell+1,\m,\sigma^{(\ell+1)},X_i)$:]   a
  candidate signature $\sigma^{(\ell+1)}=(\sigma_0,\ldots,\sigma_{2\ell+1}) $ is verified by testing if the following equalities hold:
\begin{eqnarray*}
e(\sigma_0,g) & = & e(h,\sigma_3) \cdot e(F(\m),\sigma_1) \\
e(\sigma_k,g) & = & e(\sigma_{k+1},\sigma_{2\ell+3-k}) \hbox{ for }
k \in
 \{2,\ldots,\ell\}\\
e(\sigma_{\ell+1},g)& =& e(X_i,\sigma_{\ell+2})
\end{eqnarray*}
\end{description}

\subsection{Security}

\begin{theorem} \label{Waters-proof}
The scheme with $L$ levels (and thus at most $L-1$ hops) is a secure
unidirectional PRS under the $(L-1)$-FlexDH and mCDH assumptions.
\end{theorem}
\begin{proof}
The proof is very similar to the one of theorem \ref{BLS-proof} and
replaces random oracle manipulations by  the tricks  of
\cite{BB2,Wat05}.  We prove the limited proxy and delegatee security
properties under the $(L-1)$-FlexDH assumption. The delegator
security is demonstrated under the mCDH assumption.

\paragraph{Limited proxy security.} We consider an adversary $\A_1$ with advantage $\varepsilon$. We
 describe an algorithm  $\B_1$    solving a
$(L-1)$-FlexDH instance $(A=g^a,B=g^b)$ with probability
$\varepsilon/4q_s(n+1)$, where $q_s$ is the number of signing
queries made by $\A_1$, within a comparable time.

\begin{description}
\item[System parameters:]  The    simulator $\B_1$ prepares common public parameters as
follows. It first sets $h=B=g^{b}$. The $(n+1)$-vector
$\overline{u}=(u',u_1,\ldots,u_n)$ is defined by choosing
$u'=h^{w'-\kappa \tau}\cdot g^{z'}$ and $u_i=h^{w_i} \cdot g^{z_i}$
for $i\in \{1,\ldots,n\}$ using random vectors
$(w',w_1,\ldots,w_n)\sample \Z_{\tau}^{n+1}$, $(z',z_1,\ldots,z_n)
\sample \Z_p^{n+1}$, where   $\kappa \sample \{0,\ldots,n\}$ is
randomly chosen and $\tau =2q_s$. For any message $\m=m_1\ldots m_n
\in \{0,1\}^n$, we have $$F(\m)=u' \cdot \prod_{i=1}^n u_i^{m_i}
=h^{J(\m)} g^{K(\m)}$$ for functions $J:\{0,1\}^n \rightarrow \Z$,
$K:\{0,1\}^n \rightarrow \Z_p$ respectively defined as
$J(\m)=w'+\sum_{i=1}^n w_i m_i - \kappa \tau$  and
$K(\m)=z'+\sum_{i=1}^n z_i m_i$. As in \cite{Wat05}, $\B_1$ will be
successful if $J(\m^\star)=0 $ for the message $\m^*$ of the forgery
stage whereas $J(\m)\neq 0 $ for all messages $\m\neq \m^*$ queried
for signature. Since $|J(.)|\leq \tau (n+1) \ll p$, we have
$J(\m^\star)=0  $ with non-negligible probability $O(1/\tau(n+1))$.
The adversary $\A_1$ is challenged on parameters
  $(g,h,\overline{u})$. \\ \vspace{-0.3 cm}

\item[Key generation:] for
user $i \in \{1,\ldots,N\}$, $\B_1$ defines a public key as
$X_i=A^{x_i}=g^{ax_i}$, for a random $x_i \sample \Z_p^*$, which
virtually defines user $i$'s private key as $ax_i$.  For pairs
$(i,j)$,     re-signature keys are chosen as
$R_{ij}=g^{x_i/x_j}=g^{ax_i/ax_j}$.  \\
\vspace{-0.3 cm}

\item[Signing queries:]  when a signature of signer $i$ is queried for a  message $\m$, $\B_1$   fails if
  $J(\m)=0 \bmod p$.
 Otherwise, following the technique of \cite{BB2,Wat05},
 it can construct a signature by picking $r \sample \Z_p$
and computing   \vspace{-0.2 cm} $$\sigma=
(\sigma_1,\sigma_2)=\left(
X_i^{-\frac{K(\mathsf{m})}{J(\mathsf{m})}} \cdot
F(\mathsf{m})^r,X_i^{-\frac{1}{J(\mathsf{m})}} \cdot g^r  \right).
\vspace{-0.2 cm}
 $$ which is returned to $\A_I$.  If we define $\tilde{r}=r-(ax_i)/J(\mathsf{m})$, $\sigma$ has the correct distribution
  as \vspace{-0.2 cm}
\begin{eqnarray*}
\sigma_1 = X_i^{-\frac{K(\mathsf{m})}{J(\mathsf{m})}} \cdot
F(\mathsf{m})^r
    = X_i^{-\frac{K(\mathsf{m})}{J(\mathsf{m})}} \cdot
    F(\mathsf{m})^{\tilde{r}}
    \cdot (h^{J(\mathsf{m})} \cdot g^{K(\mathsf{m})})^{ \frac{ax_i }{J(\mathsf{m})} }    = h^{ax_i} \cdot F(\mathsf{m})^{\tilde{r}} \vspace{-0.2 cm}
\end{eqnarray*}
and $\sigma_2=g^{r -(ax_i)/J(\mathsf{m})}=g^{\tilde{r}}$.
\\ \vspace{-0.3 cm}
\end{description}
After polynomially many queries, $\A_1$ comes up with a message,
that was never queried for signature for any signer, and index
$i^\star \in \{1,\ldots,N\}$ and a   forgery

 \begin{eqnarray*}
  {\sigma^{(L)}}^\star &=& (\sigma_0^\star,\ldots,\sigma_{2L-1}^\star) \\
  &=& (  h^{a
x_{i^\star}  t_1^\star \cdots t_{L-1}^\star} \cdot
F(\m)^{r^\star},g^{r^\star}, g^{ a x_{i^\star}  t_1^\star \cdots
t_{L-1}^\star}, g^{a x_{i^\star}  t_1^\star \cdots t_{L-2}^\star},
\\ & & ~ \ldots,
  g^{a
x_{i^\star}  t_1^\star },g^{t_1^\star},\cdots,g^{t_{L-1}^\star} )
\in \G^{2L}
  \end{eqnarray*}
 at level $L$. At this stage, $\B_1$
fails if $J(\m^\star)\neq 0 \bmod p$. Otherwise, if
${\sigma^{(L)}}^\star$ is valid,
$$  {\sigma_0}^\star=     h^{a x_{i^\star} t_1^\star \cdots
t_{L-1}^\star} \cdot  g^{r^\star K(\m^\star)}
$$
 which provides $\B_1$ with a valid $(2L-1)$-uple
 \begin{eqnarray*}
& & (C_1,\ldots,C_{L-1},D_1^a,\ldots,D_{L-1}^a,D_{L-1}^{ab}) \\
&=& \Big(  {\sigma_{L+1}}^\star,\ldots,{\sigma_{2L-1}}^\star,
 {{\sigma_{L}}^\star}^{1/x_{i^\star}}, \ldots ,
,{{\sigma_{2}}^\star}^{1/x_{i^\star}}, \Big( \frac{{\sigma_0}^\star
}{
{{\sigma_1}^\star}^{K(\m^\star)} } \Big)^{1/x_{i^\star}} \Big) \\
&=& \big(g^{t_1^\star},\ldots,g^{t_{L-1}^\star},g^{a
t_1^\star},\ldots, g^{a t_1^\star \cdots t_{L-1}^\star} ,g^{
t_1^\star \cdots t_{L-1}^\star ab} \big).
\end{eqnarray*} A completely
similar analysis to \cite{Wat05} shows that $J(\m^\star)=0$ with
probability $1/4q_s(n+1)$, which yields the   bound on $\B_1$'s
advantage. \\ \vspace{-0.3 cm}

\paragraph{Delegatee security.} A delegatee security adversary
$\A_2$ also implies a breach in the $(L-1)$-FlexDH assumption. The
simulator $\B_2$ is given $(A=g^a,B=g^b)$ and uses a strategy that
is completely analogous to the one of simulator $\B_1$ in the proof
of limited proxy security.

\begin{description}

\item[System parameters and public keys:] $\B_2$ prepares public parameters exactly as in the proof of
limited proxy security. The public key of the target user is defined
as $X_0=A=g^a$. The attacker $\A_2$ must be provided with private
keys for all the delegators of that user. For $i=1,\ldots,n$, other
public keys  are therefore chosen as $X_i=g^{x_i}$ for   randomly
picked $x_i \sample \Z_p^*$. The adversary $\A_2$   then receives
$\{g,h=B,\overline{u},X_0=g^a,x_1,\ldots,x_n\}$ as well as
re-signature keys $R_{ij}$ for $i\in \{0,\ldots,N\}$ and $j \in
\{1,\ldots,N\}$.
 These   are set as $R_{0j}=A^{1/x_j}=g^{a/x_j}$  and $R_{ij}=g^{x_i/x_j}$
 if $i \neq 0$.
   \\
\vspace{-0.3 cm}
\item[Signing queries:]  for all signers $i\neq 0$, $\A_2$ can generate signatures on her
own. When a signature of the target signer  is requested for a
message $\m$, $\B_2$
 proceeds as $\B_1$ did when facing the limited proxy adversary $\A_1$. It  fails if
  $J(\m)=0 \bmod p$ and can answer the query otherwise.
\\ \vspace{-0.3 cm}
\end{description}
When $\A_2$ eventually outputs a   forgery $
(\sigma_0^\star,\ldots,\sigma_{2L-1}^\star)$ at level $L$, $\B_2$
is successful if $J(\m^\star)=0$ and  extracts an admissible $(2L-1)$-uple  as $\B_1$ did. \\
\vspace{-0.3 cm}

\paragraph{Delegator security.} A delegator security adversary
$\A_3$ having advantage $\varepsilon$ after $q_s$  signing queries
is finally shown to imply an algorithm $\B_3$ to solve a problem
which is equivalent (under linear time reduction) to the mCDH
problem with probability $\varepsilon/4q_s (n+1)$. Given
$(g,A=g^a,A'=g^{1/a},B=g^b)$, this problem is to find out $g^{ab}$.
\begin{description}
\item[Public parameters and public key generation:] Again, system  parameters are  prepared  as in the proof of
limited proxy security. Namely, $\B_3$ defines $h=B=g^b$ and chooses
$u',u_1,\ldots,u_n$ so as to have $F(\m)=h^{J(\m)} \cdot g^{K(\m)}$
for some functions $J,K:\{0,1\}^n \rightarrow \Z_p$ where $J$
cancels with non-negligible probability. The public key of the
target delegator is
 set as $X_0=A=g^a$.  For
$i=1,\ldots,n$, remaining public keys  are set as $X_i=g^{x_i}$ for
a random    $x_i \sample \Z_p^*$. The adversary $\A_3$ receives
$\{g,h=B,\overline{u},X_0=g^a,x_1,\ldots,x_n\}$. This time, she is
provided with {\it all} re-signature keys (including $R_{0j}$ and
$R_{j0}$) and attempts to produce a first level forgery.  For pairs
$(i,j)$ such that $i,j \neq 0$,   $\B_3$   sets
$R_{ij}=g^{x_i/x_j}$. If $i=0$, it defines
$R_{0j}=A^{1/x_j}=g^{a/x_j}$. If $j=0$ (and thus $i \neq 0$),
$\B_3$ calculates $R_{i0}={A'}^{x_i}=g^{x_i/a}$ and hands $\{R_{ij}\}_{i,j}$ to $\A_3$. \\
\vspace{-0.3 cm}
\item[Signing queries:]   when $\A_3$ asks for a signature from the target delegator  for a
message $\m$, $\B_3$
   fails if
  $J(\m)=0 \bmod p$ and can answer the query exactly as in the proof of limited proxy security otherwise.
\\ \vspace{-0.3 cm}
\end{description}
 Eventually, $\A_3$     produces a first level
forgery $ {\sigma^{(1)}}^\star =({\sigma_1}^\star,{\sigma_2}^\star
)$ for a message $\m^\star$ that was never queried for signature. If
$J(\m^\star) \neq 0$, $\B_3$ fails. Otherwise, given that
$({\sigma_1}^\star , {\sigma_2}^\star )=(h^a \cdot g^{r
K(\m^\star)}, g^r)$, $\B_3$ finds
out $g^{ab}={\sigma_1}^\star / {{\sigma_2}^\star}^{K(\m^\star)}$. \\
\vspace{-0.3 cm}

\paragraph{External Security.}

We consider an adversary $\A_4$ with advantage $\varepsilon$. We
 describe an algorithm $\B_4$    solving a
$(L-1)$-FlexDH instance $(A=g^a,B=g^b)$ with probability
$\varepsilon/(4N(q_s+q_{rs})(n+1))$ within  comparable time, where
$q_s$ and $q_{rs}$ are is the number of signing and re-signing
queries.

\begin{description}
\item[System parameters:] The simulator $\B_4 $ prepares  common public
  parameters as in the limited proxy security proof. In addition, it
  picks at random an integer  $i^{\ast} \in \{1,\ldots,N\}$. \\
\vspace{-0.3 cm}

\item[Public key generation:] when $\A_4$ asks for the creation of
user $i \in \{1,\ldots,N\}$, $\B_4$ responds \\ \vspace{-0.3 cm}
\begin{itemize}
\item with a newly generated public key $X_i=g^{x_i}$, for a random
  $x_i \sample \Z_p^*$ if $i \neq i^{\ast}$ (s.t. $x_i$, user $i$'s
  private key, is known to the simulator);
\item with $X_{i^{\ast}} = A$ if $i = i^{\ast}$ (which virtually
  defines user $i$'s private key as $a$). \\ \vspace{-0.3 cm}
\end{itemize}

\item[Oracle queries:] $\A_4$'s queries are tackled with as
follows.\\ \vspace{-0.3 cm}
\begin{itemize}
\item[$\bullet$] {\it Signing queries}: when a signature of signer $i$
  is queried for a message $\m$,
\begin{itemize}
\item[-] $\B_4$ uses its knowledge of $x_i$ to produce the signature if
  $i \neq i^{\ast}$;
\item[-]  $\B_4$ uses the simulation from the limited proxy security proof if
  $i = i^{\ast}$ (and therefore fails if $J(\m)=0 \bmod p$).
\end{itemize}
\item[$\bullet$] {\it Re-signing queries}: for such a query on input
  $(\m,\sigma^{(\ell)},i,j)$, $\B_4$ checks if $\sigma^{(\ell)}$ is a valid $\ell^{\textrm{th}}$
  level signature on $\m$ for some $\ell \in \{1,\ldots,L-1\}$ with
  respect to the public  key $i$. If yes, $\B_4$ produces a
  first level signature on $\m$ for user $j$ (using the previous
  simulation strategy), increases its level up to $\ell+1$ (for the same
  public key) using the re-signing algorithm (with re-signature key
  simply equal to $g$) and outputs the resulting $(\ell+1)^{\textrm{th}}$ level signature. The simulation only fails  if $J(\m)=0
  \bmod p$ and $j = i^{\ast}$.
\end{itemize}
\end{description}
After polynomially many queries, $\A_4$ comes up with a message
$\m^\star$, an index $j^\star \in \{1,\ldots,N\}$ and a forgery $
{\sigma^{(L)}}^\star \in \G^{2L}$ at level $L$. Recall that
$\m^\star$ cannot have been queried to signer $j^\star $. Again,
$\B_4$ fails if $J(\m^\star)\neq 0 \bmod p$ or $j^{\star} \neq
i^{\ast}$. Otherwise, if ${\sigma^{(L)}}^\star $ is valid, $\B_4$
produces a valid $(L-1)$-FlexDH-tuple as in the limited proxy
security proof. A completely similar analysis to this proof ends up
with the announced bound on $\B_4$'s advantage.
 \qed
\end{proof}

\section{Conclusions and Open Problems}

We described the first multi-use unidirectional proxy re-signatures,
which  solves a  problem left open in 2005. The random-oracle-based
proposal also offers efficiency improvements over existing solutions
  at the first level. The other scheme
additionally happens to be the first unidirectional PRS in the
standard model. \\
 \indent Two major open problems remain. First,  it would be
interesting to see if multi-level unidirectional PRS have efficient
realizations under more classical intractability assumptions. A
perhaps more challenging task would be to find out implementations
of such primitives where the size of signatures and the verification
cost grow sub-linearly with the number of translations.

\appendix

\section{Generic hardness of $\ell$-FlexDH in bilinear groups}\label{generic}

To provide more confidence in the $\ell$-FlexDH assumption we prove a lower bound on the computational
complexity of the $\ell$-FlexDH problem for generic groups equiped with bilinear maps.
In \cite{KJP06}, Kunz-Jacques and Pointcheval define a family of computational problems that enables to study variants of the CDH problem in the generic group model. Let $\A$ be an adversary in this model and  $\varphi(X_1,\ldots,X_k,Y_1,\ldots,Y_\ell)$ be a multivariate polynomial whose coefficients might depend on $\A$'s behaviour. For values of
$x_1,\ldots,x_k$ chosen by the simulator, and knowing their encodings, the goal of $\A$ is to compute the encodings of $y_1,\ldots,y_\ell$ such that
$$\varphi(x_1,\ldots,x_k,y_1,\ldots,y_\ell)=0.$$
All elements manipulated by $\A$ are linear polynomials in $x_1,\ldots,x_k$ and some new random elements introduced through the group oracle. Let us denote $P_i$ the polynomial corresponding to $y_i$ (it is a random variable), Kunz-Jacques and Pointcheval proved the following result.
\begin{theorem}[\cite{KJP06}]
Let $d = \deg(\varphi)$ and $\mathsf{P_m}$ be an upper bound for the probability
$$\Pr[\varphi(X_1,\ldots,X_k,P_1(X_1,\ldots,X_k),\ldots,P_{\ell}(X_1,\ldots,X_k))=0]$$
Then the probability that $\A$ wins after $q_G$ queries satisfies
$$\mathsf{Succ}(q_G) \leq \mathsf{P_m} + \frac{(3q_G+k+2)}{2p} + \frac{d}{p}.$$
\end{theorem}
The choice $\phi(X_1,X_2,Y_1,\ldots,Y_{\ell+1}) = Y_{\ell+1} - X_1X_2Y_1\ldots Y_{\ell}$ implies the generic hardness of the problem $\ell$-FlexDH in groups. The purpose of this section is to prove that Kunz-Jacques and Pointcheval result also holds in generic bilinear groups and therefore that
the problem $\ell$-FlexDH is intractable in these groups.
\begin{theorem}
Let $d = \deg(\varphi)$ and $P_m$ be an upper bound for the probability
$$\Pr[\varphi(X_1,\ldots,X_k,P_1(X_1,\ldots,X_k),\ldots,P_{\ell}(X_1,\ldots,X_k))=0]$$
Then the probability that $\A$ wins after $q_G$ oracle queries to the group operations in $\G$, $\G_T$ to the bilinear map $e$  satisfies
$$\mathsf{Succ}(q_G) \leq P_m + \frac{(3q_G+k+2)}{p} + \frac{d}{p}.$$
\end{theorem}

\begin{proof} In the following $I$ and $I_T$, denote the set $\{0,\ldots,p-1\}$ and are used to represent elements of $\G$ and $\G_T$ respectively.
Following \cite{KJP06}, in the generic bilinear group model, an adversary $\A$ has access to
\begin{itemize}
\item an oracle $\mathfrak{G}$ that,
on input $(a, b, r, r') \in \Z^2 \times I^2$, answers with the representation of $ax + bx'$ in $I$, where
$r$ is the representation of $x$ and $r'$ the representation of $x'$.
\item an oracle $\mathfrak{G}_T$ that,
on input $(a, b, r, r') \in \Z^2 \times I_T^2$, answers with the representation of $ax + bx'$ in $I_T$, where
$r$ is the representation of $x$ and $r'$ the representation of $x'$.
\item an oracle $\mathfrak{E}$ that,
on input $(a, b, r, r') \in \Z^2 \times I^2$, answers with the representation of $ax + bx'$ in $I_T$, where
$r$ is the representation of $x$ and $r'$ the representation of $x'$.
\end{itemize}
The connection between
representations and elements of $\Z_p$ is managed by the simulator through two lists $\mathcal{L}$ and $\mathcal{L}_T$ of pairs
$(x,r)$ associating an element with its representation. A representation $r$ in an oracle query
input does not need to correspond to an element of $\Z_p$ in $\mathcal{L}$ or $\mathcal{L}_G$; if it does, the corresponding
element is used, otherwise a random element $x$ is drawn by the simulator in $\Z_p$ and
bound to $r$, that is, $(x, r)$ is added to $\mathcal{L}$ or $\mathcal{L}_G$. The same rule applies for the answer to
the query: if $ax + bx' = x''$ with $(x'',r'')$ in $\mathcal{L}$ or $\mathcal{L}_G$, $r''$ is answered. Otherwise, a random
representation $r''$, is chosen and $(x'',r'')$ is added
to $\mathcal{L}$ or $\mathcal{L}_G$, and the answer to the oracle query is $r''$. Overall, each oracle query adds at most 3 pairs
to $\mathcal{L}$ or $\mathcal{L}_G$.

For our problem, initially we have
$$\mathcal{L} = \{(0, r_z), (1, re), (x_1, r_1), \ldots, (x_k, r_k)\} \hbox{ and } \mathcal{L}_T = \emptyset$$
and $\A$ is given $r_z, r_e, r_1, \ldots, r_k$. $\A$'s goal is to output $r_1', \ldots, r_\ell'$ corresponding to $y_1, \ldots, y_\ell$ in $\Z_p$ that, together with the
$x_i$'s, cancel $\varphi$. The last queries of $\A$ are assumed to be of the form $\mathfrak{G}(1, 0, r_i',r_e)$. $\A$ has
won if $\varphi(x_1,\ldots,x_k,y_1,\ldots,y_\ell) = 0$ where $(y_i, r_i') \in \mathcal{L}$.

To prove the generic hardness of the problem, we consider a simulator $S'$ where random values in $\Z_p$ are replaced by formal unknowns $X_i$. Represents of elements of $\G$ (\emph{resp.} $\G_T$) correspond to linear combinations (\emph{resp.} quadratic polynomials) of these unknowns with coefficients in $\Z_p$. The simulation is similar to the one given in  \cite{KJP06} and $\A$'s goal is to output $r_1', \ldots, r_\ell'$ corresponding to linear polynomials $P_1, \ldots, P_\ell$ in $\Z_p[X_1 ,\ldots,X_,\ldots]$ that, together with the unknowns
$X_i$'s, cancel $\varphi$.

The difference between $\A$'s success probability in the two simulation occurs only if $S'$'s simulation, the representations of different polynomials (linear or quadratic) $P_1$ and $P_2$ collide in $S$'s simulation. The number of polynomials in $\mathcal{L}$ and $\mathcal{L}_T$ is upper-bounded by $3q_G+k+2$ and their degrees is at most two. Therefore, the difference appears with probability at most $(3q_G+k+2)^2/p$. As in \cite{KJP06}, the success criterion in $S'$'s simulation is stricter than in $S$'s simulation and as above the probability that $\A$ succeeds in $S$'s simulation but not in $S'$'s simulation is upper-bounded by $d/p$ (since $\varphi$ is of degree $d$ and the $P_i$'s are linear polynomial). \qed

\end{proof}


\newpage 

\begin{thebibliography}{10}

\bibitem{AF07} M.~Abe, S.~Fehr. \newblock Perfect NIZK with Adaptive Soundness. \newblock In {\em TCC'07}, {\em
LNCS} 4392, pages 118--136. Springer, 2007.


\bibitem{ADR}
J.-H. An, Y.~Dodis, and T.~Rabin.
\newblock On the security of joint signature and encryption.
\newblock In {\em Eurocrypt'02}, {\em LNCS} 2332, pages 83--107.
  Springer, 2002.

\bibitem{AFGH05} G.~Ateniese, K.~ Fu, M.~Green, S.~Hohenberger.
\newblock
Improved Proxy Re-Encryption Schemes with Applications to Secure
Distributed Storage. \newblock In {\em NDSS},  2005.

\bibitem{AFGH05bis} G.~Ateniese, K.~ Fu, M.~Green, S.~Hohenberger.
\newblock
Improved Proxy Re-Encryption Schemes with Applications to Secure
Distributed Storage. \newblock In {\em {ACM} TISSEC }, 9(1): pp.
1--30, 2006.

\bibitem{AH05} G.~Ateniese,  S.~Hohenberger. \newblock Proxy re-signatures: new
definitions, algorithms, and applications. \newblock In  {\em
{ACM} CCS'05}, pages 310--319, {ACM} Press, 2005

\bibitem{BP04} M. Bellare, A. Palacio:  The knowledge-of-exponent
assumptions and 3-round zero-knowledge protocols. Proc. of
Crypto'04, Springer LNCS Vol. 3152, 273--289 (2004)

\bibitem{BR}
M.~Bellare, P.~Rogaway.
\newblock Random oracles are practical: A paradigm for designing efficient
  protocols.
\newblock In {\em   {ACM} CCS'93}, pages 62--73, {ACM} Press, 1993.

\bibitem{BBS98} M.~Blaze, G.~Bleumer, M.~Strauss. \newblock Divertible Protocols
and Atomic Proxy Cryptography. \newblock In {\em Eurocrypt'98},
  {\em LNCS} 1403, pages 127--144, 1998.

\bibitem{BPW03} A.~Boldyreva, A.~Palacio, B.~Warinschi. \newblock Secure Proxy Signature Schemes for Delegation of Signing
Rights. \newblock Cryptology ePrint Archive: Report 2003/096,
2003.

\bibitem{BB2}
D.~Boneh, X.~Boyen.
\newblock Efficient selective-ID secure identity based encryption without
  random oracles.
\newblock In {\em Eurocrypt'04},   {\em LNCS} 3027, pp. 223--238.
  Springer, 2004.









\bibitem{BLS}
D.~Boneh, B.~Lynn,   H.~Shacham.
\newblock Short signatures from the {Weil} pairing.
\newblock In {\em Asiacrypt'01}, volume 2248 of {\em LNCS}, pages 514--532.
  Springer, 2002.





\bibitem{CH07}
R.~Canetti, S.~Hohenberger.
\newblock Chosen-Ciphertext Secure Proxy Re-Encryption.
\newblock  In {\em ACM CCS'07}. pages 185--194. {ACM} Press, 2007.



\bibitem{Cor00}
J.-S. Coron.
\newblock On the exact security of Full Domain Hash.
\newblock In {\em Crypto'00}, volume 1880 of {\em LNCS}, pages 229--235.
  Springer, 2000.


\bibitem{Dam91} I. Damg\aa rd: Towards Practical Public Key Systems Secure
Against Chosen Ciphertext Attacks. Proc. of Crypto'91, Springer
LNCS Vol. 576, 445--456 (1991)


\bibitem{Dent06} A. Dent. The Hardness of the DHK Problem in the Generic Group Model.
Cryptology ePrint Archive: report 2006/156, 2006.




\bibitem{DI}  Y.~Dodis, A.-A.~Ivan. \newblock Proxy Cryptography
Revisited. \newblock In {\em NDSS'03},  2003.







\bibitem{GA} M.~Green, G.~Ateniese. \newblock Identity-Based Proxy Re-encryption. \newblock In {\em ACNS'07},   {\em LNCS} 4521, pages
288--306. Springer, 2007.




\bibitem{Hoh06}
S.~Hohenberger.
\newblock Advances in Signatures, Encryption, and E-Cash from Bilinear Groups.
\newblock Ph.D. Thesis, MIT, May 2006.


\bibitem{HRSV07} S.~Hohenberger, G.~N.~Rothblum, a.~shelat, V.~Vaikuntanathan. \newblock Securely Obfuscating Re-encryption.
\newblock In {\em TCC'07},   {\em LNCS} 4392,  pages 233--252. Springer, 2007.











\bibitem{KJP06} S.~Kunz-Jacques, D.~Pointcheval. \newblock About the Security of MTI/C0 and MQV.
\newblock In {\em SCN'06}, {\em LNCS} 4116, pages 156--172, Springer, 2006.

\bibitem{LOSSW} S.~Lu, R.~Ostrovsky, A.~Sahai, H.~Shacham, B.~Waters.
\newblock Sequential Aggregate Signatures and Multisignatures Without Random Oracles.
\newblock   In {\em  Eurocrypt'06}, volume 4004 of {\em LNCS},
  pages 465--485, Springer, 2006.


\bibitem{MOY04} T.~Malkin, S.~Obana and M.~Yung. \newblock The Hierarchy of Key Evolving Signatures and a Characterization of Proxy Signatures.
\newblock In {\em Eurocrypt'04}, volume 3027
of {\em LNCS}, pages 306--322, Springer, 2004.


\bibitem{MUO96} M.~Mambo, K.~Usuda, E.~Okamoto. \newblock  Proxy Signatures
for Delegating Signing Operation. \newblock In {\em {ACM} CCS'96},
pages 48--57. {ACM} Press, 1996.



\bibitem{Naor} M.~Naor. \newblock  On Cryptographic Assumptions and Challenges. \newblock In {\em
Crypto'03}, {\em LNCS} 2729  pages 96--109. Springer-Verlag, 2003.





\bibitem{SCWL} J.~Shao, Z.~Cao, L.~Wang, X.~Liang. \newblock Proxy Re-Signature Schemes without Random
Oracles. \newblock In {\em Indocrypt'07}, {\em LNCS} 4859, pages
197--209. Springer, 2007.


\bibitem{Sch}
C.~P. Schnorr.
\newblock Efficient identification and signatures for smart cards.
\newblock In {\em Crypto'89}, volume 435 of {\em LNCS}, pages 239--252.
  Springer, 1989.





\bibitem{Wat05} B.~Waters. \newblock Efficient Identity-Based Encryption Without
Random Oracles. \newblock In {\em Eurocrypt'05},   {\em LNCS} 3494,
pages 114--127. Springer  2005.





\end{thebibliography}
\end{document}